# Hole Transport in Impurity Band and Valence Bands Studied in Moderately Doped GaAs:Mn Single Crystals

T. Słupiński, J. Caban and K. Moskalik

*Institute of Experimental Physics, University of Warsaw, Hoża 69, 00-681 Warsaw, Poland*
e-mail: tomslu@fuw.edu.pl

PACS: 75.50.Pp, 71.30.+h, 72.80.Ey, 61.72.-y

**Abstract**

We report on simple experiment on temperature-dependent Hall effect measurements in GaMnAs single crystalline samples with Mn composition estimated at 0.05-0.3 at.% which is slightly below the onset of ferromagnetism. Impurity band transport is visible for Mn compositions of ~0.3 at.% as a clear metallic behaviour. The results show interesting situation that the Metal-Insulator transition in GaAs:Mn occurs within the impurity band which is separated from the valence bands for Mn concentrations studied here. We also discuss on the equilibrium high temperature solubility limit of Mn in GaAs, unknown precisely in the literature.

**Introduction**

GaAs very heavily doped, or alloyed, with Mn at compositions exceeding about 1 at.% has become a prototype material for the studies of carrier-induced ferromagnetism in Mn-doped III-V semiconductors since the first observations of its ferromagnetic properties [1,2] a few years after the discovery of ferromagnetism (FM) in InMnAs [3]. Ferromagnetic properties were correlated with carrier transport in GaMnAs and still the discussion is continued about their mutual relations – e.g. recent review [4]. For instance, although it is often argued that the presence of holes in the valence bands of GaMnAs is a necessary condition for FM [5], there are experimental reports published [6] arguing that even in ferromagnetic samples with Mn composition of 5-7 at.% the Fermi level still resides in the impurity band (IB) and the transport within IB may dominate over one in the valence bands. Low hole mobility usually observed in GaMnAs and a high hole effective mass were related to the role of IB [6]. Presence of the impurity band was also detected in GaMnAs alloy for Mn composition 3.5-6.9 at.% by photoemission spectroscopy [7] (but not found in InMnAs [8]).

In this paper we present simple experiment on temperature-dependent Hall effect measurements in GaMnAs samples with Mn composition slightly below the onset of ferromagnetism (FM). We explored Mn concentrations from the range about $(1-6) \times 10^{19}$ cm$^{-3}$, or Mn/Ga compositions of about 0.05-0.3 at.%, estimated from transport. This range extends to higher values than studied in the early investigations of Mn impurity in 1960-70'ies [9-12]. Also, such Mn compositions have not got much attention in recent years because of non-ferromagnetic samples below ~1 at.%. This intermediate range is, nevertheless, interesting since electric transport already has features similar to observed at higher doping (in FM samples), e.g. the onset of IB, as we show below or as it was shown earlier [12] for GaAs:Mn samples with ~0.1 at.% of Mn.

Another motivation for these studies was the search for the limit of equilibrium solubility of Mn in GaAs at high temperatures, experimentally unknown precisely in the literature, but expected to occur at compositions about 0.1-1 at.%. Tendencies to a decomposition of random solid solution related to this limit, through a short-range ordering or clustering of Mn impurities, are usually hindered in ferromagnetic samples by the MBE growth at low substrate temperature. But still, chemical ordering effects, or various types of structural disorder, are sometimes mentioned as influencing FM properties of Mn-doped III-V semiconductors [13,14], apart from isolated point defects like Mn$_i$.

The equilibrium crystal growth (close to GaAs melting temperature) of Mn-doped GaAs with Mn compositions up to ~0.3 at.% applied in these studies has still proven fruitful, and so these Mn compositions are still below the equilibrium solubility (at such high temperature). This growth method resulted in thick single-crystalline samples facilitating studies.

Results of temperature-dependent Hall effect measurements are interpreted in terms of valence bands and impurity band transport. We show that IB transport is visible for Mn compositions of ~0.3 at.% as a clear metallic behaviour. We also discuss hole mobility observed, which could be measured due to a lack of strong anomalous Hall effect in non-FM samples. Metallic transport within IB observed, supposed to be related to the impurity band formed by Manganese states, shows interesting situation that the Metal-Insulator transition in GaAs:Mn occurs through IB separated from the valence band for Mn concentrations studied here.

**Experiment**

Single-crystalline samples were prepared in authors laboratory at University of Warsaw by the equilibrium crystal growth method, liquid encapsulated Czochralski technique, using boron oxide covering the stoichiometric GaAs melt in the p-BN crucible. Elemental Mn (5.5 N's purity) was added to the melt. Several thick (0.5-0.8 mm) homogeneous samples were prepared with different Mn compositions from the range about 0.05-0.3 at.%. Temperature-dependent transport studies (resistivity and Hall effect measurements) were performed in van der Pauw configuration at temperatures 25-400K using cryostat with He refrigerator. Magnetic field of 1000-2500 Gs was used, satisfying the low-field limit. Contacts to p-type GaAs:Mn were made using In+1%Zn alloy and were annealed for 5 minutes at 400 $^o$C.

**Results and discussion**

Fig. 1 shows Hall concentration and mobility for two GaAs:Mn samples with slightly different Mn compositions of the highest prepared. For temperatures above ~50 K (sample S) and ~75K (sample Z) thermally activated transport in the valence bands dominates. It is seen for temperatures above ~50-75K that thermal activation energy $E_a$=60-65 meV is significantly reduced comparing to the values 100-110 meV reported for Mn concentrations of $10^{16}$-$10^{17}$ cm$^{-3}$ when Mn acceptors are regarded as isolated [15,16]. This reduction may result from a formation of the impurity band for high concentrations when impurities interact. IB is regarded as centered around the energy level of isolated Mn impurity and the energy width of IB increases with a concentration of impurity centers, $\Delta E \propto N^{1/3}$, where N is the concentration of impurities. As described by Mott [17], if Fermi level resides in the impurity band, and so IB is partially filled, consequently the IB should contribute to overall transport. Returning to Fig. 1a, such IB transport may describe another distinct region of temperature dependence for sample Z visible in the results. Namely, below T~50 K a non-activated, or metallic, transport is observed, consistently with the presence of IB. Within a classification of three types of transport given by Mott [17], no hopping involving isolated Mn centers was observed for sample Z, as it is seen from an independence of Hall concentration on temperature at T=25-50K. This may be due to a big concentration of Mn centers, too big to observe hoping at temperatures as high as T=25-50K. Comparing the two samples, Z and S, we see that Mn composition in sample Z is above the onset of distinct metallic transport. Similar metallic transport was also observed earlier [12] for GaAs:Mn with ~0.1 at.% of Mn.

Before we discuss more on the impurity band, we notice how Hall mobility measured, shown in Fig. 1b, can be understood. Above T~100K, that is in the valence band transport regime, lattice (fonons) scattering is visible. Slight dependence on concentration of Mn (the difference for samples S and Z) indicates that also scattering by ionized centers contributes in this temperature range. Number of ionized centers in this range increases with temperature (more effective ionization of acceptors), contributing to a decrease of mobility, together with phonon scattering. Below T~100K a drop of mobility is observed due to a dominance of ionized impurities scattering. In the range T=50-100K mobility fairy well follows $T^{3/2}$ rule (sample S) indicating that the scattering for a fixed concentration (independent of temperature in this range) of ionized centers, coming from a compensation, takes place. For sample Z, in the range T=25-50K the Hall mobility becomes close to independent on temperature, within the accuracy of measurements. As the mobility in this range is higher than it would follow from $T^{3/2}$ rule, we argue that dominant transport at T=25-50K seems not to take place in valence bands, but in IB.

To confirm above picture of transport, we have calculated Hall concentration and Hall mobility based on 2-bands model at low magnetic field approximation, describing holes in valence bands and IB. We assumed Hall factors equal unity. As the results depend on the specific model of valence band mobility assumed, not to discuss details of this aspect, we have taken an empirical mobility for p-GaAs as discussed in a review [18] (formula 116, page R170) in the form $1/\mu = 1/(A \cdot T^{3/2} \cdot N_I^{-1}) + 1/(B \cdot T^{-2.3})$, where $N_I$ is the concentration of ionized impurities, A and B are constants, where the first term describes ionized impurity scattering and the second describes combined scattering due to phonons. Statistics of carriers was calculated assuming non-degenerate case for heavy and light holes and a single acceptor level of energy $E_a$ (simplified case, instead of more difficult IB energy range), assuming spin degeneracy of acceptor level $g = 4$, that is like in the case of nonmagnetic shallow acceptors. Effective concentration and mobility of carriers in IB was assumed as independent of temperature. With this model, the measured results could be described fairy well, as seen in Fig. 1. Concentration of Mn acceptors got from this procedure was $6.5 \times 10^{19}$ cm$^{-3}$ (sample Z) and $5.5 \times 10^{19}$ cm$^{-3}$ (sample S), which is equivalent to Mn compositions of about 0.3 at.% and 0.25 at.%, respectively.

Activation energy of valence band transport was $E_a$=62-65 meV from the fitting calculations. From the picture considered, it may be argued that this activation occurs through the transitions of holes from IB to valence bands. The impurity band (more precisely its delocalized states giving metallic conductivity) is separated from the valence bands and this separation is about the measured activation energy, ~60 meV. In other words, if the impurity band was not separated from valence bands the thermally activated valence bands hole concentration at T > 50-75K would not be observed. Based on metallic conductivity observed at T = 25-50K, the Fermi level resides in the IB. So, it may be noticed that the Metal-Insulator transition in GaAs:Mn for Mn composition studied, about 0.3 at.%, occurs solely within the IB, rather than through a merging of IB with valence bands.

The width of IB below isolated Mn acceptor level (100-110 meV) equals about 40-50 meV or more. It may be interesting to notice, applying scaling rule for the width of IB mentioned above, $\Delta E \propto N^{1/3}$, that Mn composition when IB merges with valence bands is estimated at $(110/50)^3 \cdot 0.3\% = 3.2\%$. This may be consistent with the results of Burch [6] on dominance of transport in IB, mentioned in the beginning.

**Conclusions**

In summary, we have shown that moderately doped GaAs:Mn with Mn composition up to 0.3 at.% (and might be even more) may be grown successfully at high temperature with equilibrium melt-growth methods. Temperature-dependent transport at low magnetic fields can be understood fairy well and e.g. allows to determine Mn composition. At Mn compositions about 0.3 at.% the impurity band is already formed and IB transport (metallic) can be observed at low temperatures. The results presented here are consistent with opinions that impurity band transport may dominate over valence band transport even for higher Mn compositions.

**Figure caption**

Results of temperature-dependent transport studied in two p-type GaAs:Mn samples with slightly different Mn concentrations. (a) Hall concentration measured and calculated (lines). Both, the reduction of activation energy to 60-65 meV compared to 100-110 meV known for isolated Mn acceptors, and the non-activated (metallic) transport (sample Z) below T~50K are related to the impurity band. (b) Hall mobility measured and calculated (lines). Drop of mobility below T~100K is related to a dominance of the scattering by ionized impurities. Below T~50K (sample Z) the transport occur mainly in the impurity band, with mobility close to independent of temperature. The following parameters were used for calculations: acceptor concentrations $N_A=6.5 \times 10^{19}$ cm$^{-3}$ (sample Z) and $5.5 \times 10^{19}$ cm$^{-3}$ (sample S), donor concentrations $N_D=2.2 \times 10^{17}$ cm$^{-3}$ (sample Z) and $1.9 \times 10^{17}$ cm$^{-3}$ (sample S), activation energy $E_a=62$ meV (sample Z) and 65 meV (sample S), effective hole concentration in the impurity band (IB) $6.5 \times 10^{15}$ cm$^{-3}$ (sample Z), effective mobility in IB equal 35 cm$^2$/Vs.

T. Slupinski et al.

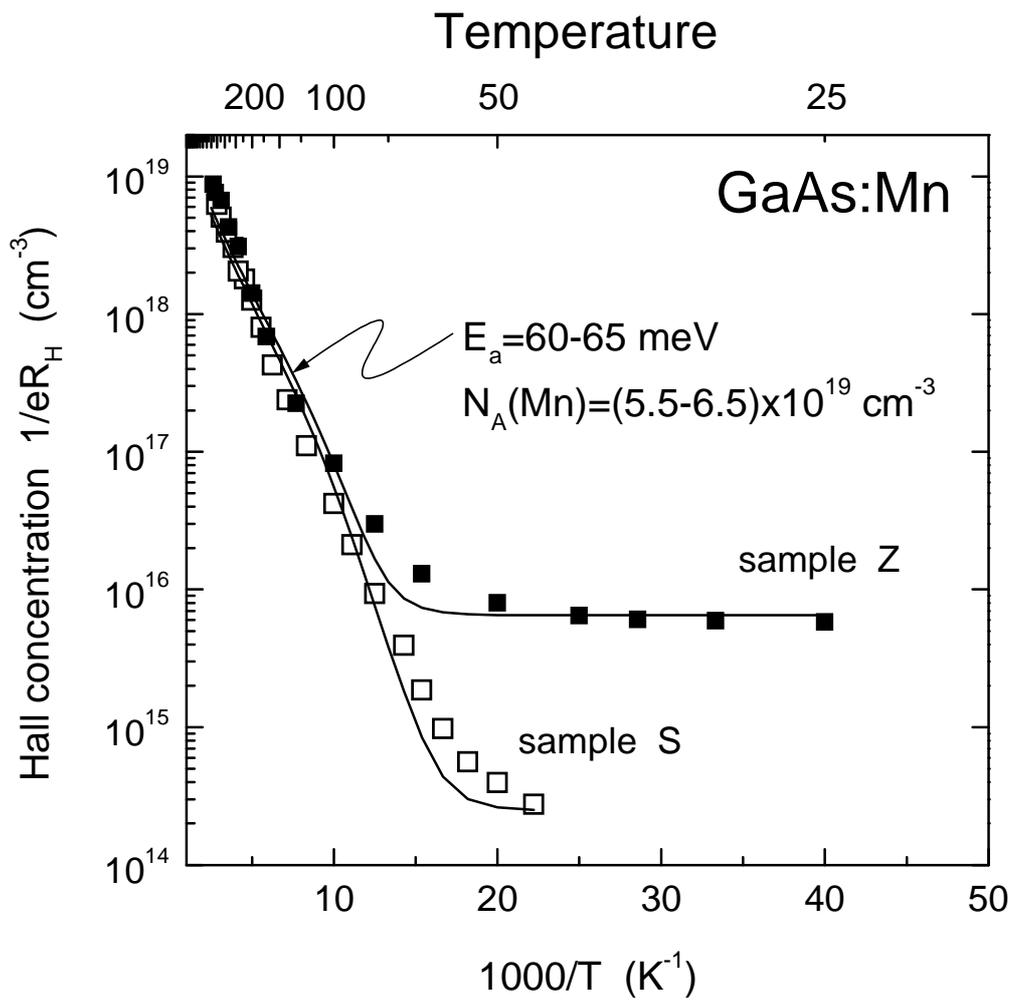

Fig. 1 a)    T. Slupinski et al.

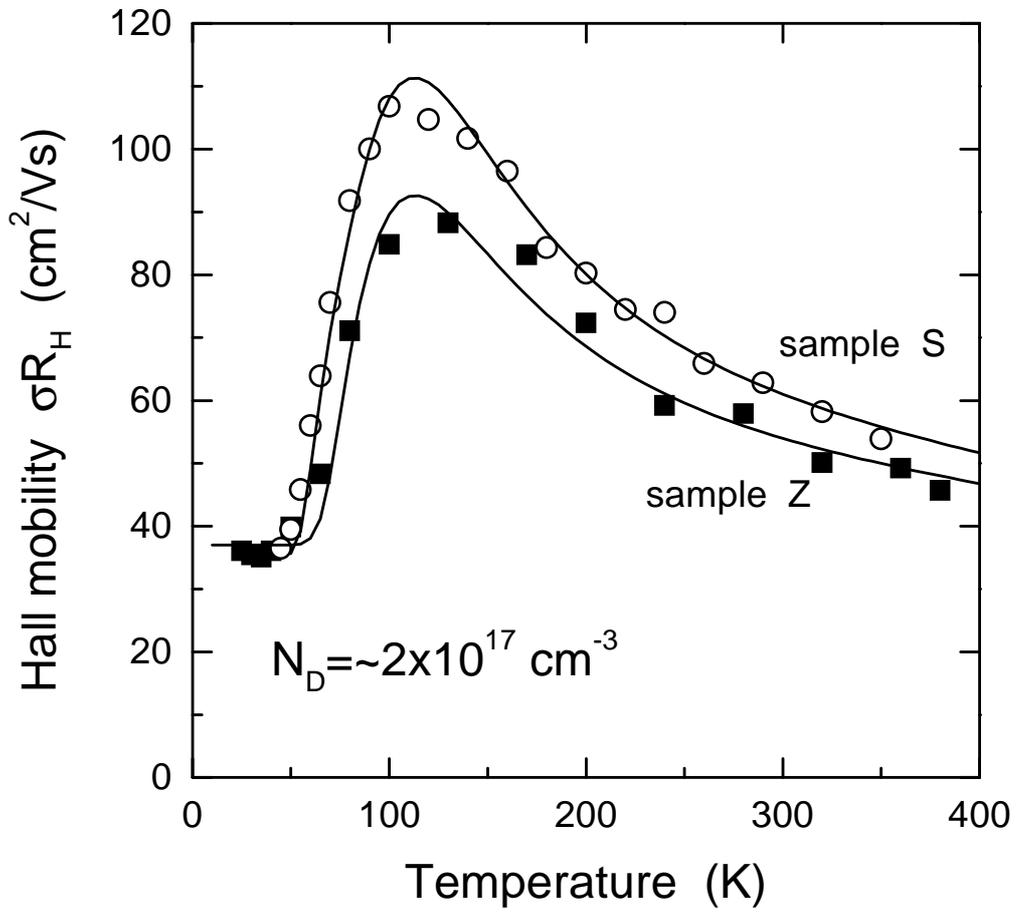

Fig. 1 b)                                                                T. Slupinski et al.